\documentclass[amsfonts, amssymb, amsmath, reprint, showkeys, nofootinbib, prd, floatfix, superscriptaddress]{revtex4-1}
\usepackage[english]{babel}
\usepackage[utf8]{inputenc}
\usepackage[colorinlistoftodos, color=green!40, prependcaption]{todonotes}
\usepackage{multirow}
\usepackage{float}
\usepackage{amsthm}
\usepackage{mathtools}
\usepackage{physics}
\usepackage{xcolor}
\usepackage{graphicx}
\usepackage[left=23mm,right=13mm,top=35mm,columnsep=15pt]{geometry} 
\usepackage{adjustbox}
\usepackage{placeins}
\usepackage[T1]{fontenc}
\usepackage{lipsum}
\usepackage{csquotes}
\usepackage[pdftex, pdftitle={Article}, pdfauthor={Author}]{hyperref} 
\newcommand{\aap}{Astron. Astrophys.}
\newcommand{\apjl}{Astrophys. J. Lett.}
\newcommand{\apjs}{Astrophys. J. Suppl.}
\newcommand{\jcap}{J. Cosmol. Astropart. Phys.}
\newcommand{\mnras}{Mon. Not. Roy. Astron. Soc.}

\defcitealias{JedamzikPogosian2020}{JP20}

\begin{document}
\title{Can small-scale baryon inhomogeneities resolve the Hubble tension?
       An investigation with ACT~DR4}

\author{Leander Thiele}
	\email{lthiele@princeton.edu}
	\affiliation{Department of Physics, Princeton University, Princeton, NJ, USA 08544}

\author{Yilun Guan}
	\affiliation{Department of Physics and Astronomy, University of Pittsburgh, Pittsburgh, PA, USA 15260}

\author{J. Colin Hill}
	\affiliation{Department of Physics, Columbia University, 538 West 120th Street, New York, NY, USA 10027}
	\affiliation{Center for Computational Astrophysics, Flatiron Institute, 162 5th Avenue, New York, NY, USA 10010}

\author{Arthur Kosowsky}
	\affiliation{Department of Physics and Astronomy, University of Pittsburgh, Pittsburgh, PA, USA 15260}

\author{David N. Spergel}
	\affiliation{Department of Astrophysical Sciences, Princeton University, Peyton Hall, Princeton, NJ, USA 08544}
	\affiliation{Center for Computational Astrophysics, Flatiron Institute, 162 5th Avenue, New York, NY, USA 10010}


\begin{abstract}
Small-scale inhomogeneities in the baryon density around recombination have been proposed as a solution
to the tension between local and global determinations of the Hubble constant.
These baryon clumping models make distinct predictions for the cosmic microwave background anisotropy power spectra on small angular scales.
We use recent data from the Atacama Cosmology Telescope to test these predictions.
No evidence for baryon clumping is found, assuming a range of parameterizations
for time-independent baryon density probability distribution functions.
The inferred Hubble constant remains in significant tension with the SH0ES measurement.
\end{abstract}


\maketitle

\section{Introduction}
\label{sec:intro}

One cornerstone of the $\Lambda$CDM standard model of cosmology is the Hubble constant $H_0$.
We may define it in two ways. In a model-agnostic way, $H_0$ measures the rate at which space
is expanding at the present day. This definition enables us to determine $H_0$ locally by measuring
the apparent speed of recession of galaxies in the Hubble flow. The other definition depends on the assumed
cosmological model: it interprets $c/H_0$ as a measurement of the angular diameter distance to a given redshift,
modulo the values of the background density parameters in the Friedmann equation. 
Measurements of $H_0$ according to this definition can be performed from the cosmic microwave background (CMB) power spectra
or the late-time matter distribution.  For both measurements,  the baryonic acoustic oscillation (BAO) feature sets a characteristic length scale.

In the correct cosmological model the two definitions of the Hubble constant
should be consistent, i.e., barring unknown systematics low- and high-redshift measurements should give consistent results.
This, however, appears not to be the case, as is illustrated in Tab.~\ref{tab:H0}.
Roughly, a $>4\sigma$ discrepancy between the low- and high-redshift inferences is observed. 

\begin{table}[b]
	\begin{tabular}{clll}
	& dataset & $H_0 [\text{km}/\text{s}/\text{Mpc}]$ & type \\
	\hline
	\hline
	\parbox[t]{2mm}{\multirow{5}{*}{\rotatebox[origin=c]{90}{local}}}
		& SH0ES \cite{Riessetal2021} & $73.2 \pm 1.3$       & SNIa+Cepheids \\
		& CSP \cite{CSP}             & $69.6 \pm 0.8 \pm 1.7$ & DEB + TRGB \\
	        & MCP \cite{MCP}             & $73.9 \pm 3.0$       & VLBI geometric \\
		& H0LiCOW \cite{H0LiCOW}     & $73.3^{+1.7}_{-1.8}$ & lensing time delay \\
		& TDCOSMO \cite{TDCOSMO}     & $67.4_{-3.2}^{+4.1}$ & lensing time delay \\
	\hline
	\parbox[t]{2mm}{\multirow{4}{*}{\rotatebox[origin=c]{90}{global}}}
		& \textit{Planck} \cite{Planck2018}          & $67.4 \pm 0.5$ & CMB \\
		& ACT \cite{ACTDR4Aiola}                     & $67.9 \pm 1.5$ & CMB \\
		& ACT+\textit{WMAP} \cite{ACTDR4Aiola,WMAP9} & $67.6 \pm 1.1$ & CMB \\
		& BOSS \cite{BOSS}                           & $68.6 \pm 1.1$ & BAO \\
	\end{tabular}
	\caption{Selection of recent $H_0$ measurements. See Sec.~\ref{sec:setup} for comment on the H0LiCOW measurement.}
	\label{tab:H0}
\end{table}

Currently, the observed discrepancy is the most severe anomaly in the $\Lambda$CDM standard model,
and hence the most promising avenue for the discovery of new physics in cosmology.
Thus, it has spurred a wide range of attempts at an explanation (so many, in fact, that we now have
dedicated reviews to keep track of them, e.g., \cite{KnoxMillea2020}).
Pre-recombination modifications include the Early Dark Energy models
\cite{KarwalKamionkowski2016,Poulinetal2019,Agrawaletal2019,Linetal2019,SaksteinTrodden2020,YePiao2020},
which however are strongly constrained by large-scale structure data \cite{Hilletal2020,Ivanovetal2020,DAmicoetal2020}.
Additional light states may also be able to accommodate higher $H_0$,
as long as their dynamics is non-trivial \cite{DEramo2018,Kreischetal2020,DasGhosh2020} (but see \cite{Brinckmannetal2020,Choudhuryetal2020}).
Post-recombination modifications such as dynamical dark energy and modified gravity
\cite{HuangWang2016,Renketal2017,Zhaoetal2017,Nunes2018,DiValentinoetal2018,Wangetal2018,Khosravietal2019,Raveri2020,Banihashemi2020,Caietal2020,DiValentinoetal2020,Dainottietal2021}
or a fifth force \cite{Desmondetal2019}
approach the tension from the opposite point of view, being not always physically well-motivated however, and often encountering problems with low-redshift expansion history data~\cite{Efstathiou2021}.

One attempted resolution of the $H_0$-tension involves modifications of $\Lambda$CDM
around the time of recombination that decrease the sound horizon $r_s$.
The peaks and troughs in the CMB power spectra are located
at wave numbers scaling as $(H_0 r_s)^{-1}$. Thus, a decrease in the sound horizon allows for higher
values of $H_0$ as inferred from the CMB.
It should be noted that this picture is somewhat simplistic, since only changing $r_s$ introduces
other tensions \cite{Jedamziketal2020}.
The model considered in this work, however, has other secondary
effects on the CMB power spectra that may be able to circumvent this constraint.

The mechanism to achieve the change in $r_s$ that we will focus on in this paper is the introduction
of $O(1)$ inhomogeneities in the baryon density around the time of recombination.
These inhomogeneities are required to be on comoving scales well below a Mpc so that they would not be
directly visible in the currently available CMB data.
Even in simple models of recombination physics it is clear that such small-scale baryon inhomogeneities
will shift the surface of last scattering to higher redshift, thus decreasing the sound horizon.
In fact, Peebles's model \cite{PeeblesPPC} contains a term $\dot n_e \propto -n_e^2$, which upon averaging gives faster recombination
in inhomogeneous plasma on account of $\langle n_e^2\rangle > \langle n_e \rangle^2$ \cite{JedamzikAbel2011}.

One attractive proposal to achieve such a scenario is the introduction of primordial magnetic fields (PMFs)
\cite{BanerjeeJedamzik2004,JedamzikAbel2011,JedamzikSaveliev2019,JedamzikPogosian2020}.
This mechanism would have the advantage that it naturally sustains the required amount of small-scale inhomogeneity through a dynamical process,
which in many other conceivable models would be entirely erased through Silk damping.
Furthermore, theoretically well-motivated PMF configurations would lead
to inhomogeneities of roughly the required amount on kpc scales.
The necessary field strengths would be such that present-day large-scale magnetic fields could
be explained with limited invocation of galactic dynamo effects.
One open question that could be answered through MHD simulations is the effect of an inverse energy cascade
from the turbulent small-scale magnetic fields and baryon inhomogeneities towards larger scales.
If such an energy flow was appreciable, it would add additional constraints.
We emphasize that this work deals with baryon inhomogeneities in an agnostic way and makes no reference to PMFs;
however, it seems difficult to generate the required inhomogeneities without invoking PMFs.
Furthermore, most other changes to recombination physics that would lead to appreciable shifts in $H_0$
are difficult to reconcile with atomic physics constraints \cite{Liuetal2020}.

While previous work -- Ref.~\cite{JedamzikPogosian2020}, hereafter \citetalias{JedamzikPogosian2020} -- restricted CMB data to the results from the \textit{Planck} satellite \cite{Planck2018},
this paper will include more recent data from the Atacama Cosmology Telescope (ACT), specifically their data release 4 (DR4) \cite{ACTDR4Choi,ACTDR4Aiola}.
Although the angular scales covered by \textit{Planck} enable tight constraints on the location of the acoustic peaks
(and thus on $H_0 r_s$), they are less constraining with regard to a secondary prediction of the inhomogeneous
models, namely the change in amplitude in the Silk damping tail \cite{JedamzikAbel2011}.
While faster recombination decreases the time photons have to diffuse (thus suppressing Silk damping),
helium recombination is also sped up which decreases the ionization fraction and therefore increases the
photon mean free path (thus enhancing Silk damping).
Generically, these two competing effects will not completely cancel; hence the inhomogeneous models have
a characteristic signature on small angular scales that is not degenerate with the shift in peak positions.
For this reason, the inclusion of small-scale ACT data provides valuable additional constraints on the proposed
resolution of the $H_0$ tension.
We may go even further and posit that the new ACT data covering scales that were not considered in \citetalias{JedamzikPogosian2020}
provide an independent test of the secondary small-scale effects predicted by this model.
If the ACT data do not show evidence of these effects, then the probability of this model resolving the Hubble tension is accordingly lessened,
compared to a hypothetical additional data set that covers the same scales as \textit{Planck}.

As the baryon inhomogeneities are on scales $10^3$ times below current resolution,
their effect is completely described by the baryon density probability distribution function (PDF).
While future MHD simulations assuming PMFs could provide constraints on the shape of this PDF,
in this work we treat it as physical input.
This has the advantage that we can treat the baryon inhomogeneities agnostically with regard to their origin,
but we will have to make certain assumptions on the shape of the PDF in order to retain a predictive model.
Future work will hopefully be able to give physically better motivated parameterizations of the PDF;
however, we believe that the range of PDFs considered in this work provides strong motivation
to believe that our results are in fact generic.

The rest of this paper is structured as follows.
In Sec.~\ref{sec:setup}, we review the three-zone model for the baryon density PDF from \citetalias{JedamzikPogosian2020} that we
adopt in this work. Futhermore, we explain our choice of likelihoods and summarize the methods used.
Sec.~\ref{sec:posteriors} contains the main results of this work, namely posteriors on $H_0$ and various
parameters of the three-zone model for a variety of likelihoods.
In Sec.~\ref{sec:bestfit} we add local $H_0$ measurements to the CMB likelihoods in order to fully excite the baryon clumping
mode and derive some intuition from the resulting best-fit models.
We conclude in Sec.~\ref{sec:conclusions}.

\section{Setup}
\label{sec:setup}

Following \citetalias{JedamzikPogosian2020}, we parameterize the baryon density PDF as a three-zone model, where each zone is described
by its density $\Delta_i = n_b^i / \langle n_b \rangle$ and its volume fraction $f_i$ ($i=1 \ldots 3$).
We can think of the comoving coordinates being split into small regions, each of which realizes one of the three zones.

We introduce the `clumping factor' $b$ as
\begin{equation}
1 + b = \sum_i f_i \Delta_i^2
\label{eq:clumping}
\end{equation}
and remind the reader of the constraints $\Sigma_i f_i = 1$ and $\Sigma_i f_i \Delta_i = 1$.
The baryon clumping models are defined by four parameters, because there are three zones,
each with density and volume fraction, and two constraints between these quantities.
For physical convenience, we choose the 4-parameter model space to be spanned by $f_2$, $\Delta_1$, $\Delta_2$,
and the clumping factor $b$ defined in Eq.~\eqref{eq:clumping}.

In this paper, for ease of calculation and comparison with \citetalias{JedamzikPogosian2020},
we consider several subsets of these models, with either one or two of the four parameters being free
and the others fixed to specific values, as listed in Tab.~\ref{tab:threezonemodels}.
These likely bracket all of the reasonable physical cases, but a more extensive analysis could consider 4-parameter
models instead.

While models M1 and M2 were chosen to facilitate direct comparison with \citetalias{JedamzikPogosian2020}, the more extended models
($f_2=\ldots$) are constructed so as to cover a relatively wide range of possible baryon density PDFs
while also being convenient computationally\footnote{Not all parameterizations are equally good, some will
only allow narrow windows in the free parameters in which the constraint system has solutions at all,
which complicates the sampling and interpretation.}.
They also constrain the variations of the baryon density to $O(1)$ excursions, consistent with the prediction
by the PMF model.

\begin{table}
	\begin{tabular}{lcccc}
	model     & $b$  & $f_2$ & $\Delta_1$ & $\Delta_2$ \\
	\hline
	$\Lambda$CDM & 0    & 1     & 0          & 1          \\
	M1           & free & 1/3   & 0.1        & 1          \\
	M2           & free & 1/3   & 0.3        & 1          \\
	$f_2=1/3$    & free & 1/3   & free       & 1          \\
	$f_2=1/2$    & free & 1/2   & free       & 1          \\
	$f_2=2/3$    & free & 2/3   & free       & 1          \\
	\end{tabular}
	\caption{
	         Versions of the three-zone model used in this work.
	         M1 and M2 are identical to the ones from \citetalias{JedamzikPogosian2020},
		 while the other three have one more degree of freedom.
		}
	\label{tab:threezonemodels}
\end{table}

We use two combinations of Boltzmann codes, recombination models, and Monte Carlo samplers to independently arrive at our
results, namely \texttt{CAMB}~\cite{CAMB} + \texttt{RECFAST}~\cite{Recfast1,Recfast2,Recfast3} + \texttt{Cobaya}~\cite{Cobaya1,Cobaya2}
and \texttt{CLASS}~\cite{CLASS1,CLASS2} + \texttt{HyRec}~\cite{HyRec} + \texttt{MontePython}~\cite{MontePython1,MontePython2}\footnote{the latter with the flag \texttt{MODEL=FULL} instead of the default \texttt{RECFAST} that ships with \texttt{CLASS}.}.
As previous work on our topic had relied on \texttt{RECFAST} which was calibrated for models close to
$\Lambda$CDM, and certainly not for $O(1)$ deviations from the average baryon density, an independent
check with the more sophisticated \texttt{HyRec} code is reassuring.
All figures in this work were produced with the second toolchain.
We run \texttt{CLASS} with the provided \texttt{cl\_permille.pre} file instead of the default precision settings,
and second Ref.~\cite{McCarthyetal2021} in recommending this as good practice,
given the high precision of current and upcoming CMB data\footnote{We performed a test with the default precision file
on the $f_2=1/2$ model. In that case, there was a negligible $\sim 0.1\sigma$ bias on $H_0$, while the shape of the $b$-posterior
was slightly but visibly altered. Thus, the precision does not seem to be an issue in this work's context, but this is not generally
true in extensions of $\Lambda$CDM constrained with post-\textit{Planck} experiments.}.
Our convergence criterion is for the Gelman-Rubin $R_\alpha$ to achieve $\max_\alpha(R_\alpha-1) < 0.05$ ($\alpha$ labels the parameters),
and we generally achieve $R_{H_0}-1 \lesssim 0.02$.

In both codes, we separately compute the recombination history $x_e^i(z)$ in each of the three zones and
then perform the average to arrive at the large-scale ionization fraction
\begin{equation}
\langle x_e(z) \rangle = \sum_i f_i \Delta_i x_e^i(z)\,.
\end{equation}

We use the following likelihoods:
\begin{itemize}
	\item \textit{Planck} 2018: high-$\ell$ TTTEEE (\texttt{lite}), low-$\ell$ TT, low-$\ell$ EE \cite{Planck2018likelihood};
	\item ACT DR4: high-$\ell$ TT, EE, TE \cite{ACTDR4Aiola,ACTDR4Choi};
	\item SMH: SH0ES \cite{SH0ES}, MCP \cite{MCP}, H0LiCOW \cite{H0LiCOW};
	\item BAO: SDSS DR7 \cite{SDSSDR7} \& BOSS DR9 \cite{BOSSDR9}, 6dFGS \cite{sixdFGS}.
\end{itemize}

We found that working with the extended \textit{Planck} likelihood (as opposed to the \texttt{lite}
version) did not give appreciable changes in the posteriors in our parameter space.

The ACT likelihood is chosen such that covariance with \textit{Planck} is negligible,
with $\ell<1800$ discarded in TT but no cuts in TE and EE \cite{ACTDR4Aiola}.

The H0LiCOW measurement used in the SMH likelihood is not the most recent one;
an updated lens modeling found substantially lower $H_0$ \cite{TDCOSMO}.
Similarly, we do not use the most recent SH0ES measurement \cite{Riessetal2021} but rather the one used in
\citetalias{JedamzikPogosian2020} from Ref.~\cite{SH0ES}.
The reason for this is that we use the SMH combination of likelihoods only for direct comparison
to \citetalias{JedamzikPogosian2020} (equal to their ``H3'') as well as in contexts in which we simply want to excite the
baryon clumping mode to a considerable extent (i.e., not in posteriors).

We assume flat priors on $\omega_b$, $\omega_\text{CDM}$, $n_s$, $A_s$, $H_0$, $\tau_\text{reio}$, $A_\text{\sf Planck}$
and the hard prior $0.9 < y_\text{p} < 1.1$ for the only ACT nuisance parameter (polarization efficiency).
When they are used, the three-zone model parameters have the wide priors $0 < b < 10$ and $0 < \Delta_1 < 0.7$
(the upper bound on $\Delta_1$ is important as it removes the spurious second posterior mode that would otherwise
be allowed by inverting the first and third zone).
We assume a fixed neutrino mass sum of $0.06\,\text{eV}$ (this helps us focus on the degeneracy
between $H_0$ and baryon clumping, rather than having to worry about the $H_0-m_\nu$ degeneracy in
addition).

\section{Results: posteriors}
\label{sec:posteriors}
  
\begin{figure}
	\includegraphics[width=0.45\textwidth]{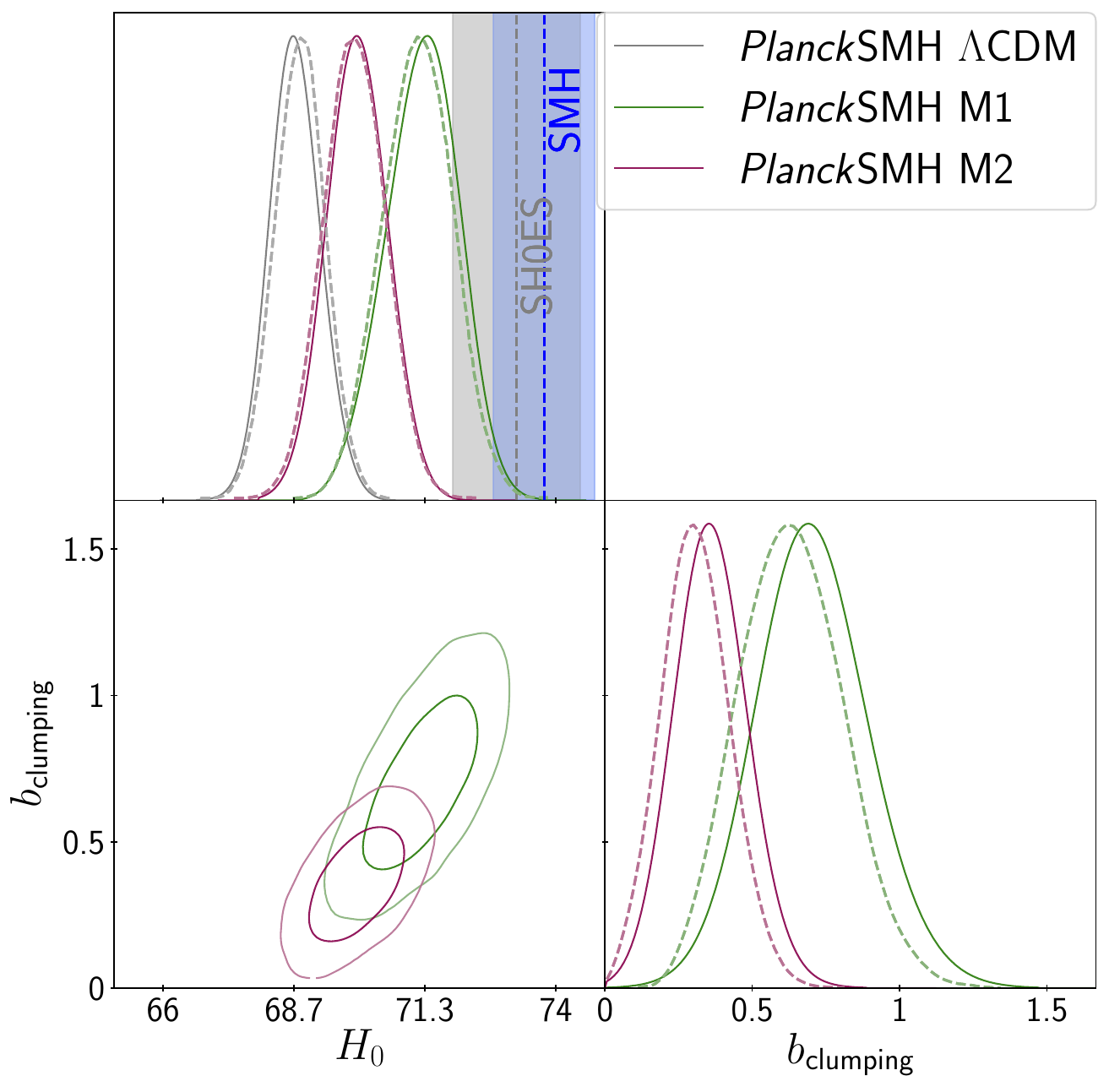}
	\caption{
		Posteriors in the same parameterizations and with the same likelihod
		as the one used in \citetalias{JedamzikPogosian2020}.
		The solid lines are our results, while the dashed lines are \citetalias{JedamzikPogosian2020}.
		Here and in all subsequent posterior plots, the grey shaded region is the
		$1\sigma$ confidence interval from the latest SH0ES measurement \cite{Riessetal2021}.
		We have also indicated the $1\sigma$ confidence interval of the SMH $H_0$ likelihood used
		in this plot and in Sec.~\ref{sec:bestfit}.
		}
	\label{fig:JP20validation}
\end{figure}

As a first step, we reproduce the results from \citetalias{JedamzikPogosian2020}, staying within their M1/M2 parameterization
of the three-zone model and working solely with the \textit{Planck}+SMH likelihood (which we adopt only in order to reproduce
previous results and assess the accuracy of \texttt{RECFAST} in the baryon clumping models).
The resulting posteriors are shown in Fig.~\ref{fig:JP20validation}.
We observe that our posteriors on $H_0$ are almost identical with those found in \citetalias{JedamzikPogosian2020}
(which are plotted as dashed lines).
There are some minor discrepancies in the posterior on the clumping parameter $b$,
which we believe are most likely explained by differences between \texttt{HyRec} and \texttt{RECFAST}\footnote{As mentioned in
Sec.~\ref{sec:setup}, we confirmed key results with the tool-chain used in \citetalias{JedamzikPogosian2020}. In that case,
we find almost perfect agreement with their results.}.

In the remainder of this section, we will not include SMH (i.e. SH0ES, MCP, and the old H0LiCOW value) in the likelihoods.
As we will see, the CMB(+BAO) likelihoods are not compatible with SMH even after allowing for baryon clumping,
which in our view renders a formally combined likelihood problematic for interpretation.

\begin{figure}
	\includegraphics[width=0.45\textwidth]{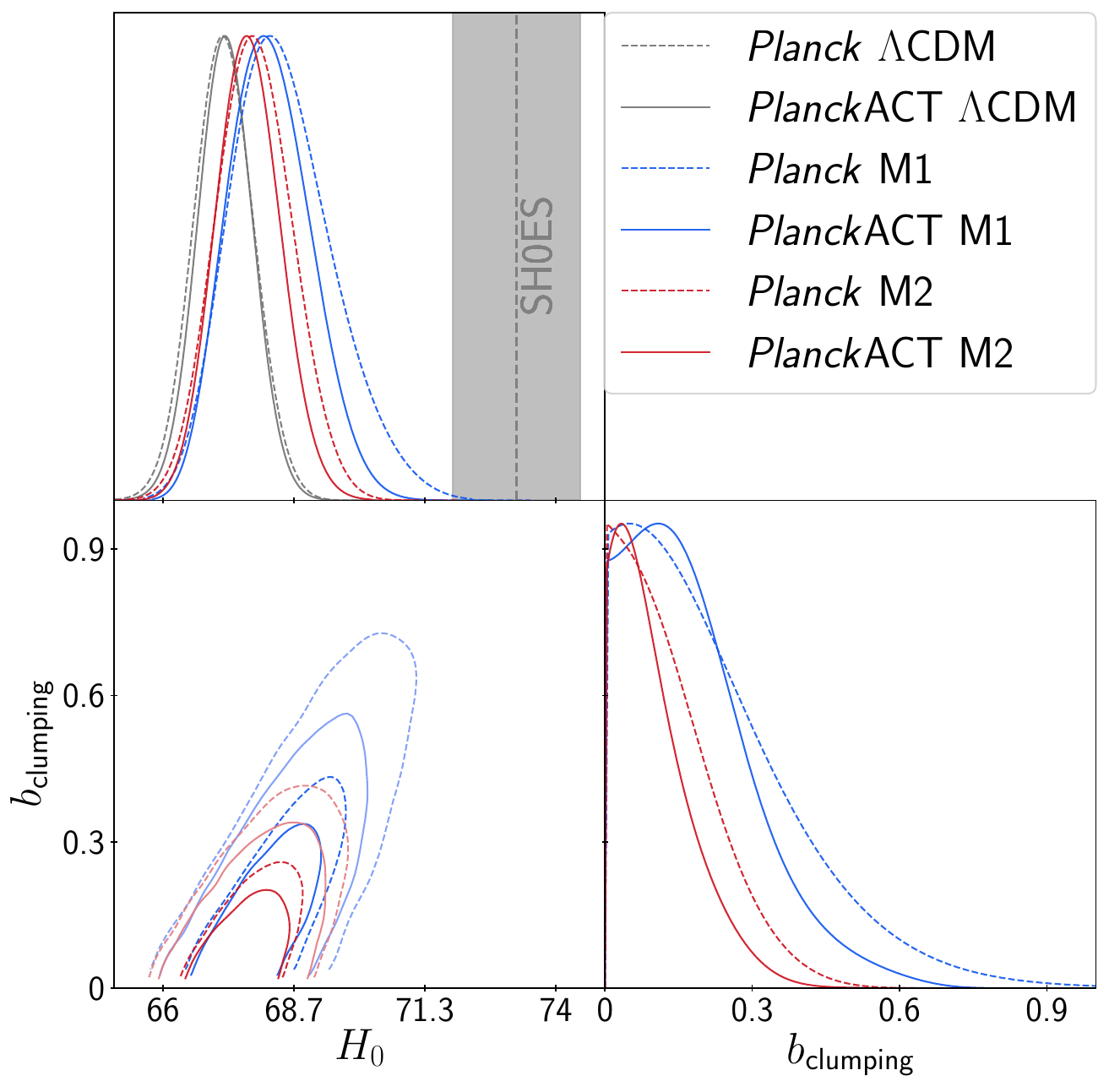}
	\caption{
	        Posteriors in the same parameterizations as in \citetalias{JedamzikPogosian2020},
		comparing \textit{Planck} alone with ACT added.
		Here and in the following posteriors, the line color indicates the model while the line style
		differentiates between different likelihoods.
	        }
	\label{fig:JP20wACT}
\end{figure}

In Fig.~\ref{fig:JP20wACT}, we compare posteriors in the \citetalias{JedamzikPogosian2020} parameterizations
fitted to \textit{Planck} and \textit{Planck}+ACT.
For these posteriors, as well as for all the others plotted in this section, central values and 1-$\sigma$ uncertainties
are listed in Tab.~\ref{tab:centralsigma} in Appendix~\ref{app:centralsigma}.
First, we observe that within $\Lambda$CDM both likelihoods yield very similar posteriors on $H_0$
(this is to be expected, as ACT DR4 itself has only about 1/3 the statistical power of \textit{Planck}, and moreover their $\Lambda$CDM constraints independently agree well~\cite{ACTDR4Aiola}).
However, in the models allowing baryon clumping, M1 and M2, the situation is different.
While the $H_0$ posterior's lower end is virtually identical for both likelihoods, the inclusion of ACT significantly
reduces the extent of the high-$H_0$ tail (due to this asymmetry, the posterior plot is more illuminating than the bare 1-$\sigma$ uncertainties).
The effect is more pronounced for model M1 which yields a less severe tension with SH0ES than M2 does.
Similarly, adding ACT reduces the $95\,\%$ upper limits on the clumping parameter $b$ (c.f. Tab.~\ref{tab:centralsigma}).
Although M1 and M2 consistently lead to slightly higher central values of $H_0$, the shift is not nearly enough to get in
reasonable agreement with the SH0ES measurement.
In summary, at least in the \citetalias{JedamzikPogosian2020} parameterizations M1/M2, adding ACT to \textit{Planck} increases
the tension with SH0ES within the baryon clumping models.

\begin{figure}
	\includegraphics[width=0.45\textwidth]{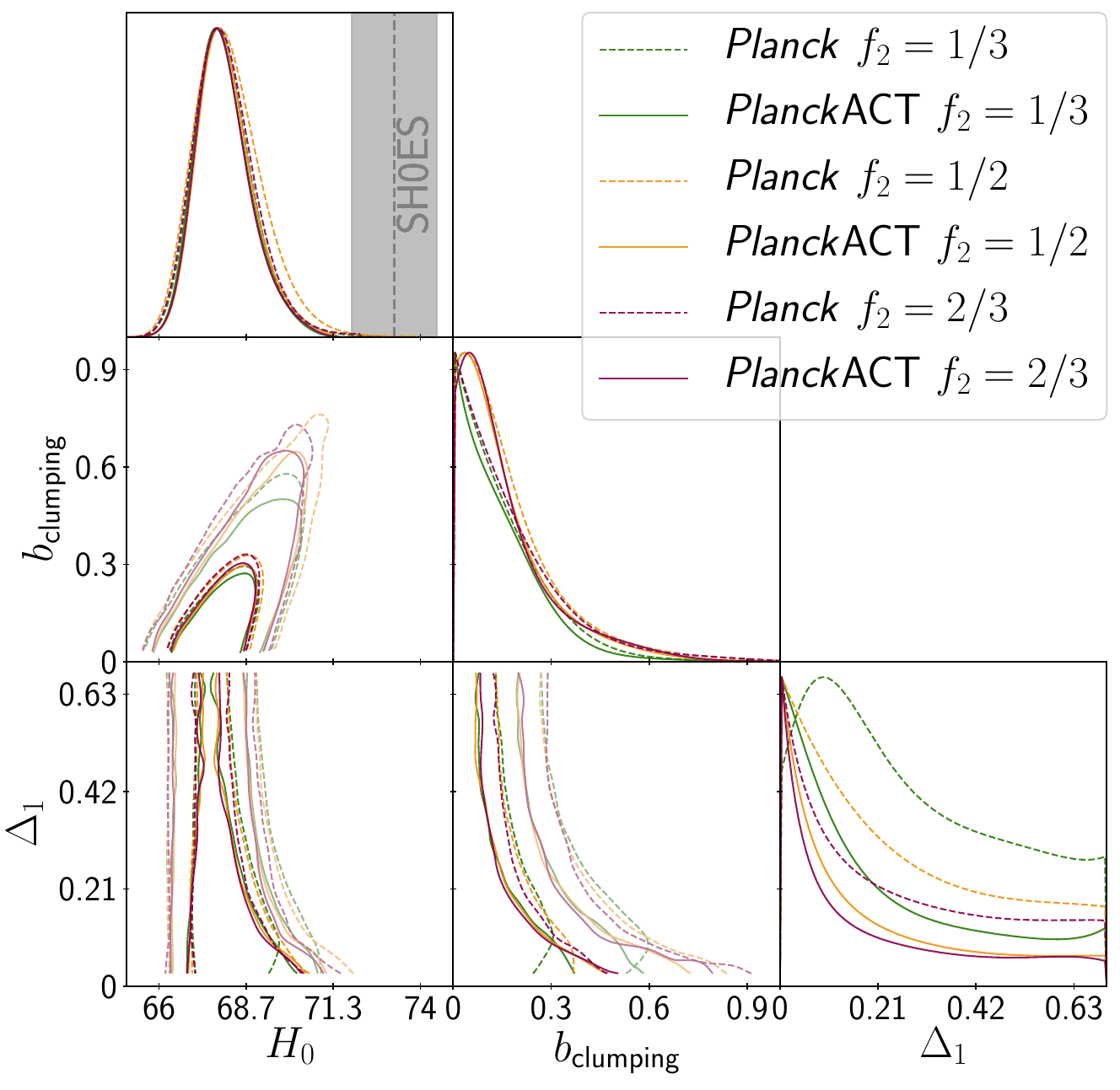}
	\caption{
		Posteriors in the parameterizations with one more degree of freedom, $\Delta_1$.
	        }
	\label{fig:free2}
\end{figure}

We proceed by adding an additional degree of freedom to the baryon density PDF, treating $\Delta_1$ as a free parameter.
The models considered here may be considered as spanning the entire range of which M1 and M2 sample only two discrete points,
while also sampling $f_2$ (which was identical for M1 and M2) at three different points.
Posteriors for the \textit{Planck} and \textit{Planck}+ACT likelihoods are shown in Fig.~\ref{fig:free2}.
We observe extremely similar posteriors in $H_0$ and $b$ for the three flavors of this extended parameterization.
Again, no detection of the clumping effect (i.e., non-zero $b$) is made.
Similarly to the posteriors shown in Fig.~\ref{fig:JP20wACT}, the addition of ACT has the most pronounced effect
in the high-$H_0$ tails, where it visibly reduces the leeway the baryon clumping models have in getting into somewhat decent
agreement with SH0ES.

The extra degree of freedom introduced in these models, $\Delta_1$, is virtually unconstrained
(for this reason, we do not include it in Tab.~\ref{tab:centralsigma}).
This may be a consequence of a poor choice of parameterization, however, it seems more likely that we are already close
to covering the entire range of CMB spectra compatible with the data and allowed by the three-zone model.
This view is supported by the fact that all three values of $f_2$ lead to extremely similar posteriors.
If this is indeed the case, the $H_0$ posteriors shown in Fig.~\ref{fig:free2} should be similar to those that could
be obtained with any reasonable baryon density PDF. As we discuss in the conclusions, time dependence may prove to be an
additional mode that could alter this picture.

We discuss a possible prior volume effect in the more extended $f_2=\ldots$ models in Appendix~\ref{app:priorvolume}.

\begin{figure}
	\includegraphics[width=0.45\textwidth]{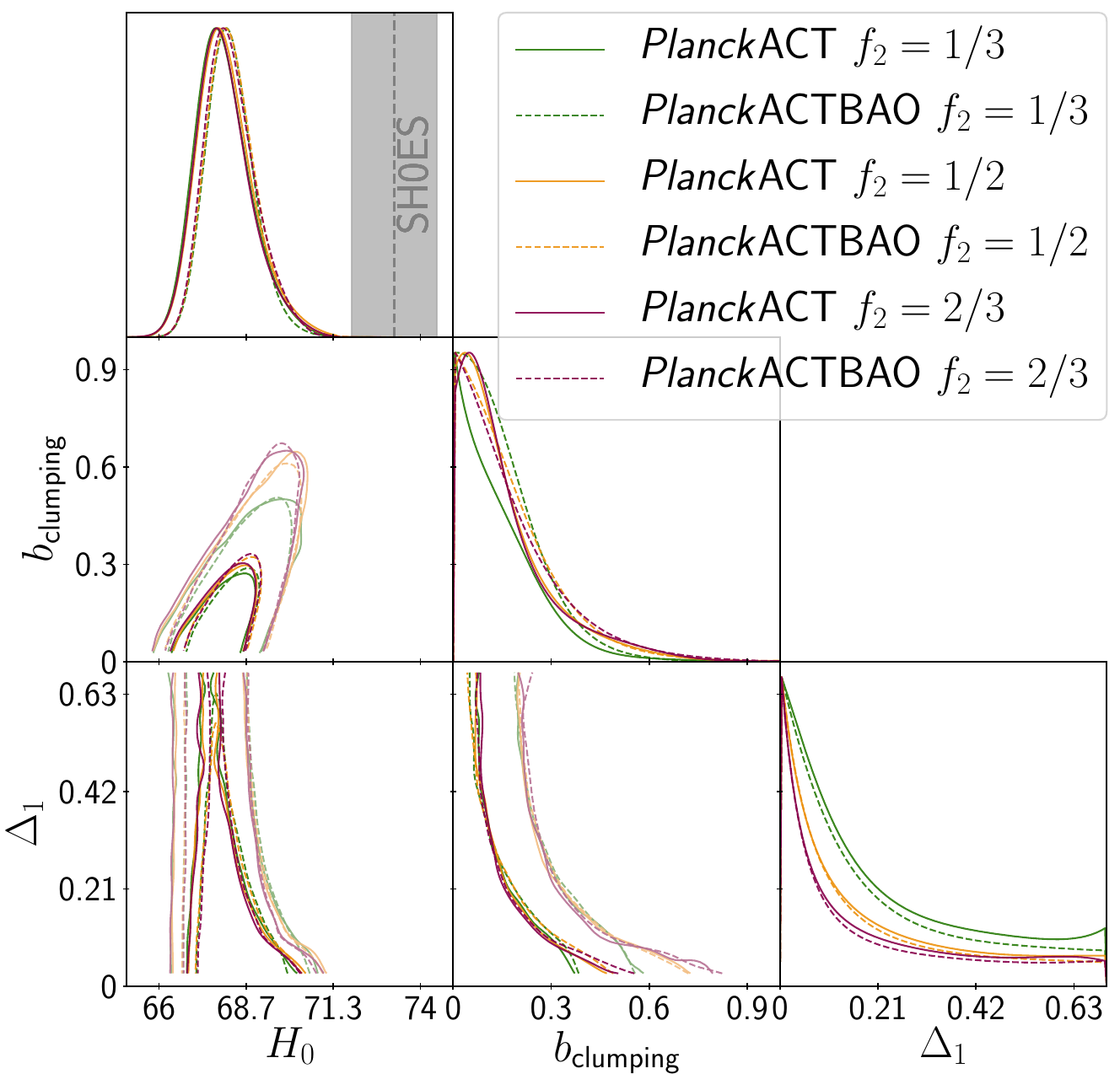}
	\caption{
		Posteriors in the extended parameterization, illustrating the effect of adding a BAO likelihood.
		}
	\label{fig:free2wBAO}
\end{figure}

As a final result of this section, we show in Fig.~\ref{fig:free2wBAO} how adding $z \lesssim 1$ BAO data affects the posteriors
in the extended models.
As for $H_0$ the BAO likelihood has comparable statistical power to the CMB data, the observation that the low-$H_0$ tail
is reduced by the addition of BAO is no surprise.
On the other hand, the high-$H_0$ tail is almost invariant under adding BAO, which indicates that the CMB data alone
are almost entirely responsible for constraining the baryon clumping modes of the model.
In fact, the posterior on $b$ is slightly widened once BAO is included.
Thus, we conclude that the BAO likelihood is not particularly useful in constraining the baryon clumping models,
at least as long as the CMB-only likelihood does not allow appreciable increases in $H_0$.

\section{Results: best-fit models}
\label{sec:bestfit}

\begin{table}
	\begin{tabular}{l|llll}
	likelihood            & \textit{Planck}+ACT & \multicolumn{3}{c}{\textit{Planck}+ACT+SMH}       \\
	model                 & $\Lambda$CDM & $\Lambda$CDM    &    M1   & $f_2=1/2$  \\
	\hline
	\hline
	$H_0$                 & 67.200  & 68.703  & 70.342  & 72.178  \\
	$100\omega_b$         & 2.2316  & 2.2561  & 2.2511  & 2.2520  \\
	$\omega_\text{cdm}$   & 0.1204  & 0.1171  & 0.1203  & 0.1246  \\
	$n_s$                 & 0.9660  & 0.9742  & 0.9652  & 0.9523  \\
	$10^9 A_s$            & 2.1086  & 2.1282  & 2.0940  & 2.0617  \\
	$\tau_\text{reio}$    & 0.0522  & 0.0687  & 0.0533  & 0.0484  \\
	\hline
	$b$                   & --      & --      & 0.4071  & 1.1822  \\
	$f_1$                 & --      & --      & 0.2865  & 0.3524  \\
	$\Delta_1$            & --      & --      & 0.1000  & 0.0048  \\
	$f_2$                 & --      & --      & 0.3333  & 0.5000  \\
	$\Delta_2$            & --      & --      & 1.0000  & 1.0000  \\
	$f_3$                 & --      & --      & 0.3801  & 0.1476  \\
	$\Delta_3$            & --      & --      & 1.6784  & 3.3757  \\
	\hline
	$A_\text{\sf Planck}$ & 1.0043  & 1.0042  & 1.0030  & 1.0018  \\
	$y_\text{p}$          & 1.0012  & 1.0016  & 1.0026  & 1.0016  \\
	\end{tabular}
	\caption{
	         Best-fit models used in Sec.~\ref{sec:bestfit}.
		 The first group of parameters are $\Lambda$CDM, the second group
		 describes the three-zone model (not all of them are independent),
		 and the third group are the nuisance parameters for our likelihoods.
		}
	\label{tab:bestfit}
\end{table}

In order to gain some further intuition, in this section we attempt to fully excite the baryon clumping mode
and explore its consequences for the CMB spectra.
In order to do this, we now add the previously discussed SMH likelihood to \textit{Planck}+ACT.
For this combined likelihood, we compute three best-fit models, listed in Tab.~\ref{tab:bestfit}
(the model in the first column, \textit{Planck}+ACT $\Lambda$CDM, was not fitted to SMH and serves as the reference point).
The models including baryon clumping, M1 and $f_1=1/2$, were chosen because they are the ones that had the highest
central values for $H_0$ in the posteriors discussed in the previous section.
We emphasize that it can be difficult to infer general trends from individual best-fit models,
so the results in this section should be taken as starting points to develop intuition.
The main results of this work have been presented in the previous section.

\begin{figure}
	\includegraphics[width=0.45\textwidth]{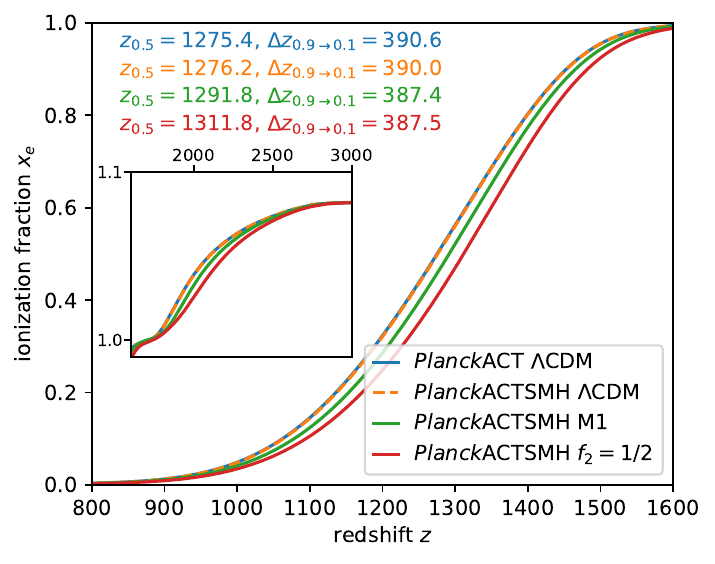}
	\caption{
		The ionization fraction as a function of redshift for the best-fit models
		listed in Tab.~\ref{tab:bestfit}.
	        }
	\label{fig:xez}
\end{figure}

For each of the best-fit models, we compute the average ionization fraction $x_e(z)$, shown in Fig.~\ref{fig:xez}.
In the figure, we also list $z_{0.5}$ and $\Delta z_{0.9 \rightarrow 0.1}$ as proxies for the location and width
of the last scattering surface.
First, we observe that without inclusion of baryon inhomogeneities the recombination history is essentially fixed.
As expected, once the additional degrees of freedom in the baryon density PDF are introduced,
recombination is pushed to higher redshift.
If we write $A\,\delta \log \Delta z_{0.9 \rightarrow 0.1} = \delta \log z_{0.5}$ (where $\delta$ is with respect to
the \textit{Planck}+ACT $\Lambda$CDM reference model), we find $A = -1.60$ and $A = -3.53$ for M1 and $f_2=1/2$ respectively.
This is, to order of magnitude, consistent with $A = -2$, the expectation from the argument that the duration of recombination
should scale as $v_\text{rms}^{-1} \propto T^{-1/2} \propto z^{-1/2}$.
However, it is interesting to note that both M1 and $f_2=1/2$ yield almost identical values for $\Delta z_{0.9 \rightarrow 0.1}$.

\renewcommand{\thempfootnote}{\arabic{mpfootnote}}
\begin{figure}
	\includegraphics[width=0.45\textwidth]{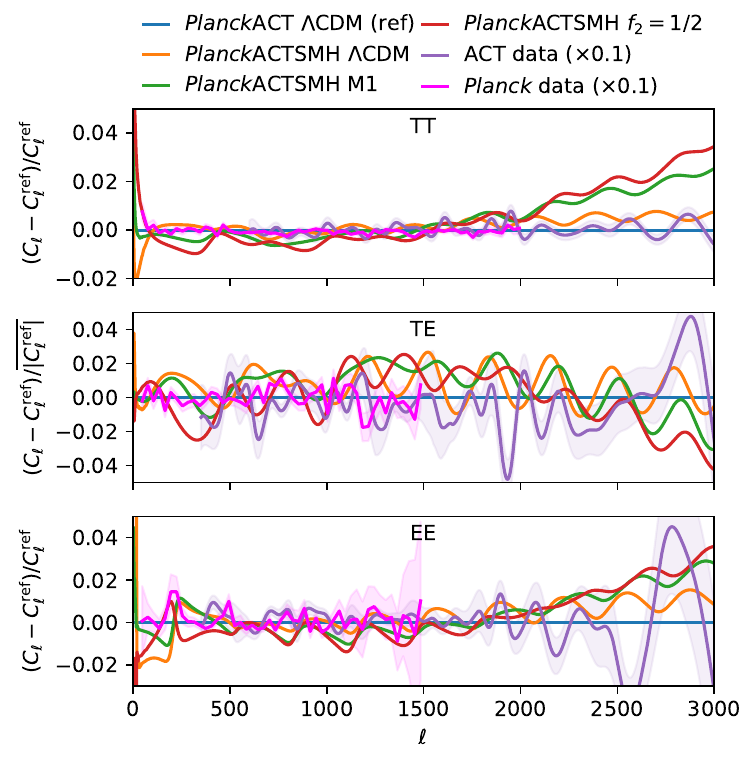}
	\caption[bla]{
		Fractional changes in CMB power spectra for the best-fit models listed in Tab.~\ref{tab:bestfit},
		with respect to the \textit{Planck}+ACT $\Lambda$CDM baseline.
		In the second panel (TE), because the data have zero crossings, we computed
		the normalizing factor $\overline{|C_\ell^\text{ref}|}$ as a convolution of the
		absolute value of the reference TE power spectrum with a Gaussian filter of width $\sigma = 100$ (in $\ell$).
		This gives a useful normalization without artifacts from the roots.
		Note that for clarity the experimental data curves for ACT and \textit{Planck} have been rescaled
		by the given factors.
		While the \textit{Planck} data points are relatively close to what we use in our likelihood
		(except for binning and $\ell$-cuts to improve clarity, in particular for TE and EE),
		the ACT data points are for illustration purposes only and were generated
		as explained in the footnote.\protect\footnotemark[5]
	        }
	\label{fig:spectra}
\end{figure}

\footnotetext{The ACT DR4 data only contain band powers, which are related
		to the underlying CMB power spectra by $C_b = W_{b\ell} C_\ell$, where $W_{b\ell}$ is an $m \times n$ matrix
		with $m < n$. Thus, it is not possible to uniquely infer the $C_\ell$ from the band powers.
		In order to generate power spectra that can be directly compared to \textit{Planck} and the theory models,
		we first find some solution $C_\ell^0$ of the above equation (which is, of course, completely unphysical).
		Then, writing $K_{\ell\alpha}$ for the $n \times (n-m)$ matrix whose columns are a basis for the kernel of $W_{b\ell}$, we can
		generate additional solutions $C_\ell(a_\alpha) = C_\ell^0 + K_{\ell\alpha}a_\alpha$, where $a_\alpha$
		is an arbitrary vector.
		We now find the $a_\alpha$ that minimizes $||\Lambda_{\ell'\ell}C_\ell(a_\alpha)||$,
		where $\Lambda_{\ell'\ell}$ implements a discrete second derivative.
		We perform this procedure for both the `deep' and `wide' parts of the ACT data set,
		associating approximate error bars with the resulting $C_\ell$ according to the normalized diagonal
		elements of the covariance matrix of the $C_b$.
		Finally, we take the weighted average of the `deep' and `wide' $C_\ell$.}

Next, we consider the CMB power spectra arising from the listed best-fit models, shown in Fig.~\ref{fig:spectra},
where we compare them to \textit{Planck} and ACT data.
As already alluded to in Sec.~\ref{sec:intro}, the models with baryon inhomogeneities show significant deviations from
$\Lambda$CDM in the high $\ell$ part of the spectra, particularly for TT and EE.
These deviations are partly caused by shifts in other parameters, as is already evident from the orange curve
depicting $\Lambda$CDM fit to \textit{Planck}+ACT+SMH.
However, in simple experiments where we put baryon inhomogeneities of magnitude comparable to the models shown here
into otherwise unchanged $\Lambda$CDM cosmologies, similar trends occur (the same happens if $H_0$ is increased so as to
neutralize the shift in peak positions).
In particular, we always observe an \emph{increase} in TT power at high-$\ell$ once the baryon inhomogeneities are
introduced.
Thus, we conclude that the effect of Silk damping is in fact reduced in the inhomogeneous models.
This is contrary to previous statements in the literature, which presumably rested on numerical experiments
made with much higher amounts of inhomogeneity.

It is interesting to disentangle how the suppression in the $\ell < 1000$ part of the EE spectrum is generated.
The ratio of EE to TT for $\ell \gg 1$ can be approximated as \cite{ZaldarriagaHarari1995}
\begin{equation}
\frac{C_\ell^{EE}}{C_\ell^{TT}} = \left( \frac{\eta_* \Delta\eta_*}{r^2 (1+R)}  \right)^2 B_\ell\,,
\label{eq:EETT}
\end{equation}
where $\eta_*$ and $\Delta\eta_*$ are the location and thickness of the last scattering surface in conformal time respectively,
$r$ is its distance from us, $R = 3\rho_b/4\rho_\gamma$, and $B_\ell$ is a geometric factor independent of cosmology.
Comparing, for example, the best-fit models \textit{Planck}+ACT $\Lambda$CDM and \textit{Planck}+ACT+SMH $f_2=1/2$, we find that the differences in the various
quantities entering Eq.~\eqref{eq:EETT} are all $\sim 2\ldots4\,\%$.
On aggregate, however, the cancellation gives a $\sim 1\,\%$ effect.

The ACT data do not seem to show a preference for the general trends in the spectra introduced by the baryon inhomogeneities.

\begin{table}
	\begin{tabular}{l|rrr}
	likelihood                    & \multicolumn{3}{c}{\textit{Planck}+ACT+SMH} \\
	model                         & $\Lambda$CDM  & M1    & $f_2=1/2$       \\
	\hline
	\hline
	\textit{Planck} high-$\ell$ TTTEEE     &  5.16 &  3.35 &  1.82           \\
	\textit{Planck} low-$\ell$ TT          & -1.24 &  0.31 &  2.86           \\
	\textit{Planck} low-$\ell$ EE          &  1.53 &  0.02 & -0.18           \\
	\hline
	ACT TTTEEE                    &  1.12 &  2.13 & -0.20           \\
	ACT TT                        & -1.82 & -3.57 & -3.72           \\
	ACT TE                        &  3.07 &  5.52 &  3.29           \\
	ACT EE                        & -0.33 &  1.08 &  2.17           \\
	ACT high-$\ell$ TT            & -1.66 & -3.09 & -2.77           \\
	ACT high-$\ell$ TT+TEEE       &  1.29 &  3.46 &  1.79           \\
	\hline
	SH0ES                         & -9.06 & -16.39& -21.43          \\
	MCP                           & -1.99 & -3.58 & -4.66           \\
	H0LiCOW                       & -5.25 & -9.29 & -11.74          \\
	\hline
	CMB likelihood                &  6.74 &  7.14 &  6.29           \\
	SMH likelihood                & -16.30& -29.26& -37.83          \\
	\end{tabular}
	\caption{
		Values of $\Delta\chi^2$ for three best-fit models from Tab.~\ref{tab:bestfit},
		compared to the \textit{Planck}+ACT $\Lambda$CDM reference.
		The final two lines show the combined \textit{Planck}+ACT and SMH likelihoods used in this work.
	        }
	\label{tab:DeltaChiSq}
\end{table}

Finally, we present the changes in $\chi^2$ relative to the \textit{Planck}+ACT $\Lambda$CDM baseline for the other three best-fit models,
listed in Tab.~\ref{tab:DeltaChiSq}.
The model M1 increases $\chi^2$ by about equal amounts in \textit{Planck} and ACT,
while the model $f_2 = 1/2$ with one extra degree of freedom in the baryon density PDF is comparably more
disfavoured by \textit{Planck} than by ACT.
ACT TT actually appears to favour the baryon clumping models, which holds true for both the entire $\ell$-range
covered by ACT and $\ell>1800$, the part used in our likelihoods.
On the other hand, ACT TE and, to a lesser extent, EE, disfavour baryon clumping.
This may be explained by the inability of those models to keep the last scattering surface's thickness fixed (c.f. Fig.~\ref{fig:xez}).
Another possibility is simply that the steeper slopes in the polarization spectra translate into higher sensitivity to $H_0$.
It should be mentioned that there have been hints of inconsistencies between ACT TE and other CMB data~\cite{ACTDR4Aiola}.
As no mechanism has been proposed that could explain these trends, and the tensions are still relatively mild, we do not see
this work as the appropriate place to investigate them further.

\section{Conclusions}
\label{sec:conclusions}

The introduction of inhomogeneities in the baryon density around the time of recombination
is an attractive proposal to resolve the Hubble tension. While we do not constrain ourselves
to a particular mechanism that is supposed to generate the required small-scale power,
PMFs appear to be the most natural candidate.

A key prediction of the inhomogeneous models is, besides the geometric shift in acoustic peak
positions, a change in the high-$\ell$ part of the CMB power spectra due to different amount of Silk damping.
We have argued that the addition of small-scale ACT data to the previously considered \textit{Planck} likelihood
constitutes a powerful way to test this prediction.
The higher statistical power of the thus enlarged likelihood also enables us to extend the parameter space;
specifically, we are able to add an additional degree of freedom to the baryon density PDF.

Fitting the baryon clumping models to CMB-only likelihoods, we have shown that the inclusion of ACT
narrows the posterior for the clumping parameter $b$ around zero compared to \textit{Planck} alone.
Likewise, although ACT does not appreciably alter the low-$H_0$ tails in the posteriors, it significantly
shrinks the high-$H_0$ tails, indicating that the added small-scale information disfavours the introduction
of baryon clumping.
The CMB data still leave room for mild baryon clumping and some upward shift in $H_0$, thus somewhat relieving
the Hubble tension, but values as high as the SH0ES 1-$\sigma$ lower bound are strongly disfavoured.

Although the absence of theoretical predictions for the baryon density PDF restricts us to relatively
simple parameterizations, we have argued that there is evidence that the posteriors on $H_0$ presented
in this work are in fact generic for any time-independent baryon density PDF.
An intriguing possibility that could evade the constraints presented would be the introduction of time dependence.
Assuming the existence of PMFs, the decrease in magnetic field strength during recombination
would lead to a homogenization of the baryon distribution (the relaxation time for kpc-scale baryon clumps is
considerably smaller than recombination's duration). It is possible that, in the large-scale average,
this would lead to a faster onset of recombination while keeping the thickness of the last scattering surface
approximately fixed, perhaps leading to a better fit with the polarization data.

Besides our primary results, we have also shown that while the previously used \texttt{RECFAST} recombination code
is probably sufficiently accurate for the modifications considered in this work, there are some systematic differences
to the more recent \texttt{HyRec} which we recommend should be used going forward.
Furthermore, we have shown that, contrary to previous claims, baryon inhomogeneities of the amount necessary to
appreciably shift $H_0$ do not generically lead to enhanced Silk damping.

We have found no evidence that the addition of small-scale ACT data to the likelihood favours the introduction
of baryon inhomogeneities around recombination over $\Lambda$CDM.
Looking forward, the increased small-scale CMB information from future ACT releases, SPT-3G~\cite{Benson2014}, Simons Observatory~\cite{SO2019}, and CMB-S4~\cite{CMBS42019} will
provide more constraining power. As long as no currently unknown systematics in ACT are discovered, however,
smaller error bars will only serve to exclude the class of time-independent three-zone models considered in this paper with higher significance.
Parallel to this improvement in data quality, theoretical advancements in constructing viable models of the baryon
density PDF, for example through MHD simulations, will be essential in increasing the predictivity of the theory.

Modifications to \texttt{CLASS} used in this work are publicly available \href{https://github.com/leanderthiele/class\_baryon\_inhom}{here}.

\begin{acknowledgments}
We thank Karsten Jedamzik and Levon Pogosian for useful discussions during the final stages
of this work.
JCH and DNS thank the Simons Foundation for support.
LT thanks Zack Li for help with the ACT DR4 likelihood.
\end{acknowledgments}

\bibliography{main}{}

\onecolumngrid
\appendix

\clearpage

\section{Assessment of prior volume effects}
\label{app:priorvolume}

The more extended parameterizations considered in this paper, $f_2=\ldots$, exhibit a pathology
in that $f_1=0$ is equivalent to $\Lambda$CDM regardless of the value of $\Delta_1$.
Thus, there may be a prior volume effect in these models that underestimates the posterior for
non-vanishing baryon clumping (and thus increased $H_0$).
We assess this effect for the example of the $f_2=1/2$ model; for other choices of $f_2$ the conclusions
are similar.
Our methods are similar to those used in Ref.~\cite{Ivanovetal2020}.

\begin{figure}
	\begin{minipage}{0.45\textwidth}
	\includegraphics[width=\textwidth]{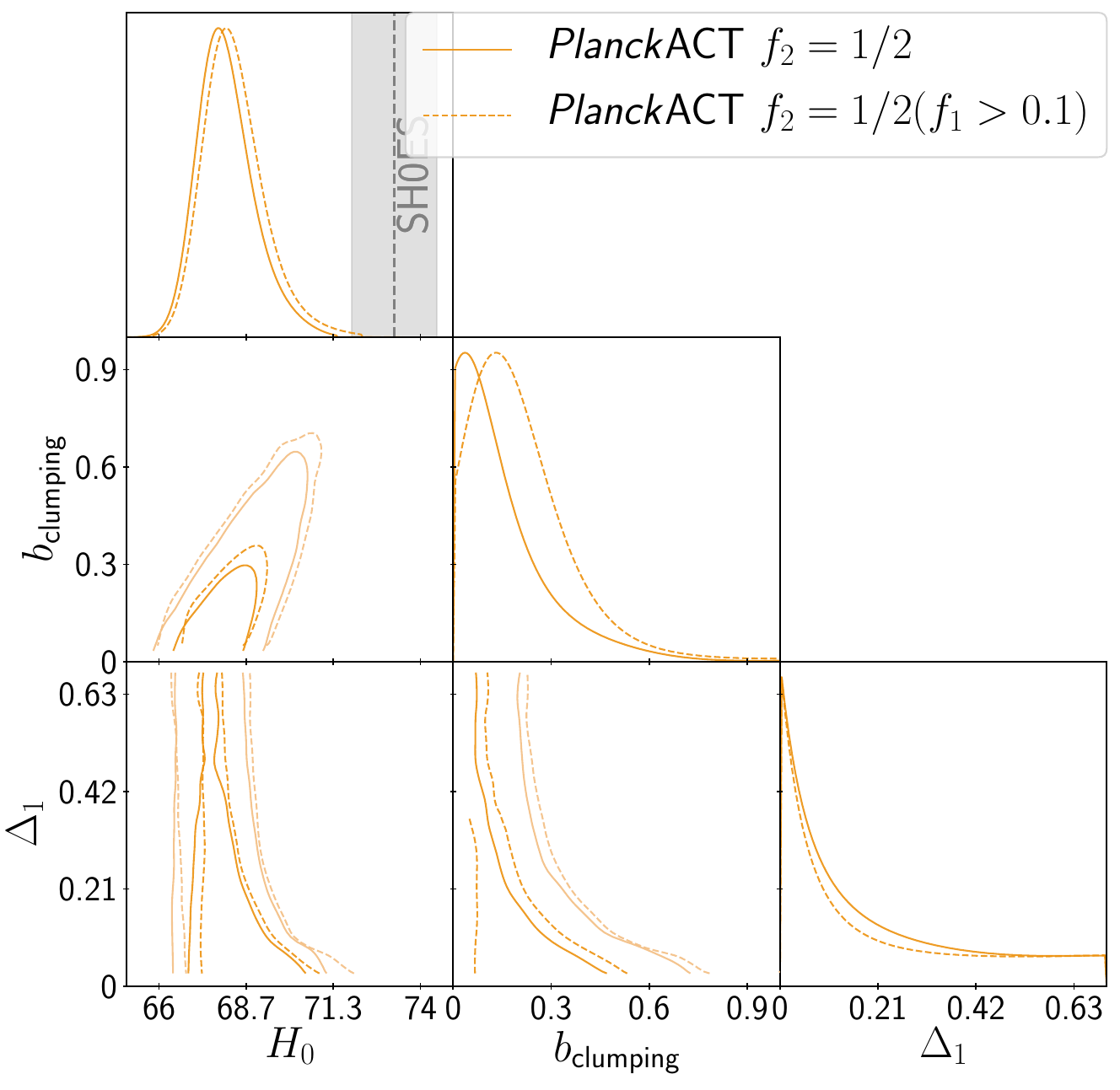}
	\caption{
		Comparison between the posterior in the unconstrained $f_2=1/2$ model,
		same as in Fig.~\ref{fig:free2}, and the one with the constraint $f_1>0.1$.
		The latter should substantially reduce the prior volume effect arising from the degeneracy
		with $\Lambda$CDM at $f_1=0$.
		In the dashed posterior, the non-vanishing support at $b=0$
		is an artifact of \texttt{MontePython}'s posterior smoothing routine.
		}
	\label{fig:f1constraint}
	\end{minipage}%
	\hspace{0.05\textwidth}
	\begin{minipage}{0.45\textwidth}
	\includegraphics[width=\textwidth]{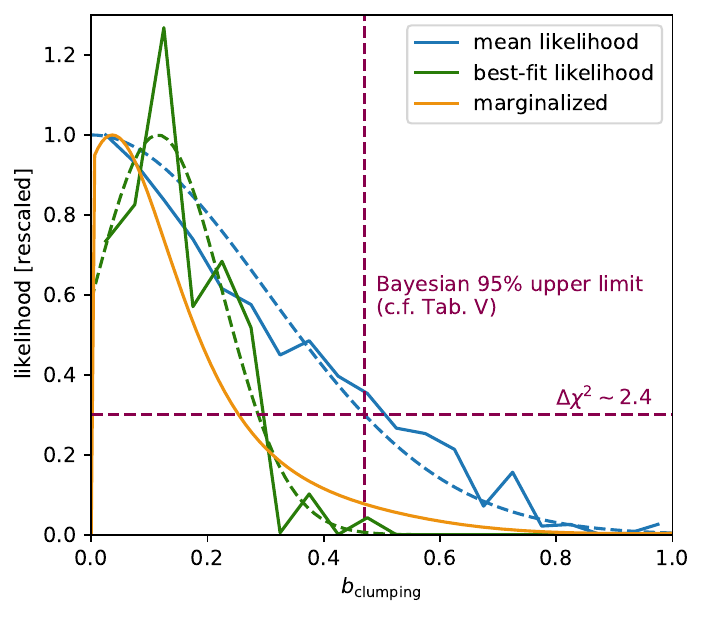}
	\caption{
		Likelihood profiles as a function of $b$ for the $f_2=1/2$ Markov chain.
		The `mean' and `best-fit' lines are computed as described in the text.
		The orange line is the same marginalized likelihood that was shown in Fig.~\ref{fig:free2}.
		The blue dashed line is a Gaussian fit with only the standard deviation varied,
		while the green dashed line is a 3-parameter Gaussian fit.
		The green solid line was rescaled so that the Gaussian fit attains its maximum at one.
		These fits should only be treated as visual guides.
		}
	\label{fig:lprofile}
	\end{minipage}
\end{figure}

First, we consider changing the model such that a constraint $f_1>0.1$ is enforced, thereby avoiding $\Lambda$CDM entirely.
The resulting posteriors are shown in Fig.~\ref{fig:f1constraint}.
It is expected that the posteriors for $b$ and $H_0$ move upward to some extent since the model,
by construction, contains at least some baryon clumping now.
The shift in the central value of $H_0$ is about $\sigma/3$ and we do not observe a substantial
enlargement of the high-$H_0$ tail.

Second, we compute a frequentist mean likelihood profile in $b$, which is the exponential of the average log-likelihood
extracted from the Markov chain in 20 $b$-bins between 0 and 1, shown as the solid blue line in Fig.~\ref{fig:lprofile}.
We observe that this profile is somewhat broader than the posterior shown in Fig.~\ref{fig:free2}.
Furthermore, at the $95\,\%$ upper limit on $b$ from Tab.~\ref{tab:centralsigma}, the averaged likelihood
is reduced by about $\Delta\chi^2\sim 2.4$. This value would correspond to only an $\sim 88\,\%$ upper bound.

We also compute the more common likelihood profile in which we minimize the likelihood at a range of values of $b$,
shown as the solid green line in Fig.~\ref{fig:lprofile}.
In this case, we observe that the maximum is at non-zero $b$, but then the likelihood profile drops off more steeply than
the marginalized Bayesian posterior.
For a discussion of the relative merits of the two ways to compute a likelihood profile, see Ref.~\cite{Ivanovetal2020}.

We conclude that we observe evidence for a mild prior volume effect. However, the posteriors
in Fig.~\ref{fig:f1constraint} indicate that our main conclusions regarding $H_0$ are only marginally
affected.

\clearpage

\section{Central values and uncertainties}
\label{app:centralsigma}

\begin{table}[h]
	\setlength\tabcolsep{2mm}
	\begin{tabular}{ll|cccccccc}
	likelihood & model & $H_0 \, [{\rm km/s/Mpc}]$ & $100\omega_b$ & $\omega_\text{cdm}$ & $n_s$ & $10^9 A_s$ & $\tau_\text{reio}$ & $b$ \\
	\hline
	\hline
	\multirow{6}{*}{\textit{Planck}}
	& $\Lambda$CDM & $67.24 \pm 0.66$ & 2.23(2) & 0.120(1) & 0.964(5) & 2.10(4) & 0.054(9) & -- \\
	& M1           & $68.45 \pm 1.08$ & 2.24(2) & 0.122(2) & 0.960(5) & 2.10(4) & 0.053(8) & <0.57 \\
	& M2           & $67.86 \pm 0.83$ & 2.24(2) & 0.121(2) & 0.964(5) & 2.10(4) & 0.055(9) & <0.32 \\
	& $f_2=1/3$    & $68.04 \pm 0.97$ & 2.24(2) & 0.121(2) & 0.963(6) & 2.10(4) & 0.054(9) & <0.42 \\
	& $f_2=1/2$    & $68.12 \pm 1.04$ & 2.24(2) & 0.121(2) & 0.963(6) & 2.10(4) & 0.054(9) & <0.53 \\
	& $f_2=2/3$    & $68.03 \pm 0.96$ & 2.24(2) & 0.121(2) & 0.964(6) & 2.10(4) & 0.054(9) & <0.52 \\
	\hline
	\multirow{6}{*}{\textit{Planck}+ACT}
	& $\Lambda$CDM & $67.26 \pm 0.60$ & 2.23(1) & 0.120(1) & 0.967(4) & 2.12(4) & 0.055(8) & -- \\
	& M1           & $68.18 \pm 0.87$ & 2.23(1) & 0.121(2) & 0.962(5) & 2.11(4) & 0.053(8) & <0.42 \\
	& M2           & $67.74 \pm 0.71$ & 2.24(1) & 0.121(1) & 0.967(4) & 2.12(4) & 0.054(8) & <0.26 \\
	& $f_2=1/3$    & $67.99 \pm 0.86$ & 2.23(1) & 0.121(2) & 0.964(6) & 2.11(4) & 0.054(8) & <0.39 \\
	& $f_2=1/2$    & $68.06 \pm 0.88$ & 2.23(1) & 0.121(2) & 0.964(6) & 2.11(4) & 0.054(8) & <0.47 \\
	& $f_2=2/3$    & $68.01 \pm 0.89$ & 2.24(2) & 0.121(2) & 0.965(5) & 2.11(4) & 0.054(8) & <0.49 \\
	\hline
	\multirow{3}{*}{\textit{Planck}+ACT+BAO}
	& $f_2=1/3$    & $68.18 \pm 0.77$ & 2.24(2) & 0.121(2) & 0.964(6) & 2.11(4) & 0.054(8) & <0.38 \\
	& $f_2=1/2$    & $68.24 \pm 0.83$ & 2.24(1) & 0.121(2) & 0.964(6) & 2.11(4) & 0.054(8) & <0.46 \\
	& $f_2=2/3$    & $68.22 \pm 0.79$ & 2.24(1) & 0.121(2) & 0.966(5) & 2.11(4) & 0.054(9) & <0.50 \\
	\end{tabular}
	\caption{
		Central values and 1-$\sigma$ errors for the posteriors in this section.
	        For conciseness, for $\Lambda$CDM parameters except $H_0$, the numbers in parentheses
		are the 1-$\sigma$ uncertainties on the last digit.
		Since we do not observe a detection of the clumping effect, for the clumping parameter $b$
		$95\,\%$ upper limits are quoted.
	        }
	\label{tab:centralsigma}
\end{table}

\end{document}